\documentclass[jounral]{IEEEtran}
\usepackage{subfigure}

\usepackage{caption}
\usepackage{graphicx}
\usepackage{subfig}
\usepackage{amssymb}

\ifCLASSINFOpdf

\else

\fi

\usepackage{amsmath}

\newtheorem{theorem}{\textbf{Theorem}}

\newtheorem{assumption}{\textbf{Assumption}}
\newtheorem{definition}{\textbf{Definition}}
\newtheorem{remark}{\textbf{Remark}}

\usepackage{algorithm}

\usepackage{algorithmic}

\begin{document}
\title{Joint Coverage and Power Control in Highly Dynamic and Massive UAV Networks: An Aggregative Game-theoretic Learning Approach }

\author{Zhuoying~Li,
        Pan~Zhou,~\IEEEmembership{Member,~IEEE,}
        Yanru~Zhang,~\IEEEmembership{Member,~IEEE,}
				~and Lin~Gao,~\IEEEmembership{Senior~Member,~IEEE}
\thanks{Z.~Li is with the School of Optical and Electronic Information, Huazhong University of Science and Technology, Wuhan 430074, China (e-mail: zhuoyingli@hust.edu.cn).}
\thanks{P. Zhou is with the School of Electronic Information and Communications, Huazhong University of Science and Technology, Wuhan 430074, China (email: panzhou@hust.edu.cn).}
\thanks{Y.~Zhang is with the School of Computer Science and Engineering, University of Electronic Science and Technology of China, Chengdu  611731, China (email: yanruzhang@uestc.edu.cn).}
\thanks{L.~Gao is with the Department of Electronic and Information Engineering, Harbin Institute of Technology, Shenzhen 518055, China (email: gaol@hit.edu.cn).}}

\markboth{IEEE Transactions on Wireless Communication,~Vol.~X, No.~X, August~2019}%
{Shell \MakeLowercase{\textit{et al.}}: Bare Demo of IEEEtran.cls for IEEE Journals}

\maketitle

\begin{abstract}
Unmanned aerial vehicles (UAV) ad-hoc network is a significant contingency plan for communication after a natural disaster, such as typhoon and earthquake. To achieve efficient and rapid networks deployment, we employ noncooperative game theory and amended binary log-linear algorithm (BLLA) seeking for the Nash equilibrium which achieves the optimal network performance. We not only take channel overlap and power control into account but also consider coverage and the complexity of interference. However, extensive UAV game theoretical models show limitations in post-disaster scenarios which require large-scale UAV network deployments. Besides, the highly dynamic post-disaster scenarios cause strategies updating constraint and strategy-deciding error on UAV ad-hoc networks. To handle these problems, we employ \emph{aggregative game} which could capture and cover those characteristics. Moreover, we propose a novel \emph{synchronous} payoff-based binary log-linear learning algorithm (SPBLLA) to lessen information exchange and reduce time consumption. Ultimately, the experiments indicate that, under the same strategy-deciding error rate, SPBLLA’s learning rate is manifestly faster than that of the revised BLLA. Hence, the new model and algorithm are more suitable and promising for large-scale highly dynamic scenarios.
\end{abstract}


\IEEEpeerreviewmaketitle

\section{Introduction}
\subsection{Background and Motivation}

\IEEEPARstart {C}{atastrophic} natural and man-made disasters, such as earthquakes, typhoons, and wars, usually involve great loss of life and/or properties, historical interests in vast areas. Though sometimes unavoidable, the loss of life and property can be effectively reduced if proper disaster management has been implemented. Since telecommunication infrastructure can be vital for search and rescue missions after disasters \cite{SahS}, when communication networks are damaged by natural disasters, rapid deployment to provide wireless coverage for ground users is quite essential for disaster management \cite{DohP}. Considering that repairing communication infrastructures takes a long time, building vehicle relay networks was a preferable solution during the critical first 72 hours \cite{AkbV}. In such relay networks, vehicles are distributed in the post-disaster areas and act as communication infrastructures, supporting mission-critical communication. However, vehicles are likely barricaded by damaged roads, torrential rivers, and precipices, etc, in the vast areas. Faced with those challenges, unmanned aerial vehicles (UAV) ad-hoc networks become a good alternative for emergency response due to its numerous advantages such as quick deployment and resilience in large-scale harsh conditions \cite{TunG}\cite{SahO}.

\par To investigate UAV networks, novel network models should jointly consider power control and altitude for practicability. Energy consumption, SNR and coverage size are key points to decide the performance of a UAV network \cite{PirA}. Respectively, power control determines the signal to energy consumption and noise ratio (SNR) of UAV; altitude decides the number of users that can be supported \cite{TroA}, and it also determines the minimum value of SNR. It is because the higher altitude a UAV is, the more users it can support, and the higher SNR it requires. Therefore, power control and altitude are two essential factors. There have been extensive researches building models focusing on various network influence factors. For example, the work in \cite{CheJ} established a system model with channels and time slots selections. Authors of \cite{oldLLA} constructed a coverage model which considered each agent's coverage size on a network graph. However, such models usually consider only one specific characteristic of networks but ignore systems' multiplicity, which would bring great loss in practice. Since UAVs will consume too much power to improve SNR or to increase coverage size. Even though UAV systems in 3D scenario with multi-factors of coverage and charging strategies have been studied by \cite{TroA}, it overlooks power control which means that UAVs might wast lots of energy. To sum up, in terms of UAV ad-hoc networks in post-disaster scenarios, power control and altitude which determine energy consumption, SNR, and coverage size ought to be considered to make the model credible \cite{TuD}.

\par In post-disaster scenarios, a great many of UAVs are required to support users \cite{TunG}. Therefore, we propose aggregative game theory into such scenarios and permit UAV to learn in the constrained strategy sets. Because the aggregative game can integrate the impact of all other UAVs on one UAV, it reduces the complexity of receiving information and reduces the data processing capacity of UAVs. For instance, in a conventional game applied a scenario with N UAVs, it needs to analyze N strategies which decide noise and coverage sizes from each other individual UAV. However aggregative game only needs to process the integrated noise and coverage sizes of all other UAVs. Such an advantage is more obvious when the number of UAVs is extremely large since figuring out each others' strategies is unrealistic \cite{CheJ}. In terms of constrained strategy sets, due to environmental factors such as violent winds \cite{FanJ} and tempestuous rainstorms, the action set of UAVs has a restriction that cannot switch rapidly between extreme high power or elevate altitude to low ones, but only levels adjacent to them \cite{TunG2}. For instance, the power can change from $1mW$ to $1.5mW$ in the first time slot and from $1.5mW$ to $2mW$ in the next one, but it cannot alter it directly from $1mW$ to $2mW$. Therefore, the aggregative game with constrained sets is an ideal model for post-disaster scenarios.

\par A new algorithm which can learn from previous experiences is required, and the algorithm with faster learning speed is more desirable. Existing algorithms' learning method is learning by prediction. It means that UAV knows current strategies with corresponding payoff and it can randomly select another strategy and calculate its payoff. And then UAV compares two payoffs. If the payoff of new strategy is larger, the current strategy will be replaced by the new strategy; if the current payoff strategy is large, it will remain in the current strategy. However, under highly dynamic scenarios, complicate network conditions make UAVs hard to calculate their expected payoffs but only learn from previous experiences \cite{FanJ}. In this situation, UAV only can learn from previous experience. UAV merely knows the current and the last strategy with the corresponding payoff, and it can only learn from these. In this case, if the two strategies are different, UAV chooses the strategy with the larger payoff as the current strategy in the next iteration; if the two strategies are the same, a new strategy is randomly selected as the current strategy of the next iteration. To sum up, the difference between the existing and required algorithms is that the existing algorithm calculates payoff for a new strategy (equivalent to prediction) and chooses it based on prediction; while required algorithm can be available when UAV only knows strategy's payoff if strategy has been selected, and then UAV can decide whether to stick to the current strategy or return to the past strategy by comparing their payoffs. Therefore, an algorithm which can learn from previous experiences is required.

\par The learning rate of the extant algorithm is also not desirable \cite{BluL}. Recently, a new fast algorithm called binary log-linear learning algorithm (BLLA) has been proposed by \cite{MarJ}. However, in this algorithm, only one UAV is allowed to change strategy in one iteration based on current game state, and then another UAV changes strategy in the next iteration based on the new game state. It means that UAVs are not permitted to update strategies at the same time. Besides, to determine which UAV to update strategy, the coordinating process will occupy plenty of channel capacities and require more time between two iterations \cite{LinL}. If the algorithm can learn synchronously, more than one UAV can update strategies based on the current game state in one iteration. Thus, the algorithm can be more efficient. To sum up, synchronous update algorithms which can learn from previous experiences are desirable, but only a little research investigated on it.

\subsection{Contribution}
We establish a multi-factor system model based on large-scale UAV networks in highly dynamic post-disaster scenarios. Considering the limitations in existing algorithms, we devise a novel algorithm which is capable of updating strategies simultaneously to fit the highly dynamic environments. The main contributions of this paper are as follows:
\begin{itemize}
\item We design a model which jointly considers multiple factors such as coverage and power control in multi-channel scenario. The model with more network influence factors ensures its reliability.
\item We propose a novel UAV ad-hoc network model with the aggregative game which is compatible with the large-scale highly dynamic environments, in which several influences are coupled together. In the aggregative game, the interference from other UAVs can be regarded as the integral influence, which makes the model more practical and efficient.
\item We revise BLLA and construct a novel payoff-based binary log-linear learning algorithm (PBLLA). PBLLA outperforms BLLA in the sense that it is capable of learning previous utilities and strategies to fit for post-disaster scenarios.
\item We propose the synchronous payoff-based binary log-linear learning algorithm (SPBLLA) which has the following properties: 1) SPBLLA can learn with restricted information; 2) In certain conditions, SPBLLA approaches NE with constrained strategies sets; 3) SPBLLA allows UAVs to update strategies synchronously, which significantly speeds up the learning rate; 4) SPBLLA avoids system disorder and ensures synchronous learning, which is a main breakthrough.
\end{itemize}

\par We organize this paper as follows. In section II, we introduce the related works. In section III, we first introduce the UAV's power control in the multi-channel communication and coverage problems, then form a system model in highly dynamic scenarios. Moreover, in section IV, we formulate our work as an aggregative game and prove the existence of the NE. In section V, we propose the two algorithms for approaching the NE. Section VI presents the simulation results and discussions. Ultimately, section VII gives a conclusion of the whole study.

\section{Related Work}
The literature mainly lies in four areas, we will present the studies of UAV communication network, channel reuse in the communication network, game theoretical approaches, and the implementation algorithms sequentially.

\subsection{UAV Communication Network}
\par With the rapid commercialization of UAVs, a lot of research has emerged in this field \cite{KaiP}. To efficiently deploy UAVs, studies have been made to find out UAV distribution on network graph \cite{oldLLA} and a graphical model has been proposed for channels reuse \cite{XuY}. The resource allocation of channel and time is also a hot area to study \cite{CheJ}. As mentioned previously, those models are impractical for post-disaster scenarios since they fail to jointly consider power control and coverage which are coupled factors.

\subsection{Channel Reuse}
\par Typical wireless protocol 802.11b/g only provides limited channels for users, which is far more than enough for high-quality communication services \cite{LiH}. To reduce the load in central system, making use of distributed available resources in networks turns out to be an ideal solution. Underlay Device-to-Device (D2D) communication is considered as one of the crucial technologies for cellular spectrum reuse for user devices in communication networks \cite{DomS}. The advantage of D2D communication that allows end users to operate on licensed channels through power control sheds light on how interference management would work in UAV ad-hoc networks \cite{KosJ}.

\subsection{Game Theoretical Approaches}
\par Game theory provides an efficient tool for the cooperation through resource allocation and sharing \cite{PanZ1}\cite{PanZ2}. A computation offloading game has been designed in order to balance the UAV's tradeoff between execution time and energy consumption \cite{MesM}. A sub-modular game is adopted in the scheduling of beaconing periods for the purpose of less energy consumption \cite{KouS}. Sedjelmaci et al. applied the Bayesian game-theoretic methodology in UAV's intrusion detection and attacker ejection \cite{SedH}. However, most existing models focus on common scenarios with less number of UAVs, which are not compatible with large-scale scenarios with large numbers of UAVs \cite{Bekm}. Aggregative game is a characteristic game model which treats other agents' strategies as a whole influence, thus avoids overwhelming strategies information from every single agent \cite{JenM}\cite{CorR}. Inspired by this, our model is built upon the aggregative game theory which suits for large-scale scenarios.

\subsection{Cooperative Algorithms}
\par Compared with other algorithms, novel algorithm SPBLLA has more advantages in learning rate. Various algorithms have been employed in the UAV networks in search of the optimal channel selection \cite{WuD}\cite{ChaU}, such as stochastic learning algorithm \cite{XuY2}. The most widely seen algorithm--LLA is an ideal method for NE approaching \cite{oldLLA}\cite{AliMS}. The BLLA has been employed by \cite{RahS}, which is modified from LLA to update strategies in each iteration to converge to the NE. However, only a single agent is allowed to alter strategies in one iteration. In large-scale scenarios, more iterations are required, which makes BLLA inefficient. It is obvious that more UAVs altering strategies in one iteration would be more efficient. To achieve it, the works in \cite{HasM} and \cite{HasM2} have provided a novel synchronous algorithm. However, there exist superabundant restrictions that make the algorithm impractical in most scenarios. Compared with the formers, SPBLLA has fewer constraints and can achieve synchronous operation, which can significantly improve the computational efficiency.

\par In summary, our work differs significantly from each of the above-mentioned works, and other literatures in UAV ad-hoc networks. As far as we know, our proposed algorithm is capable of learning previous utilities and strategies, achieve NE with restricted information and constrained strategies sets, and update strategies synchronously, which significantly speed up the learning rate.

\setlength{\abovecaptionskip}{0cm} 
\setlength{\belowcaptionskip}{-0.25cm} 

\begin{figure}
\centering
\includegraphics[width=\linewidth]{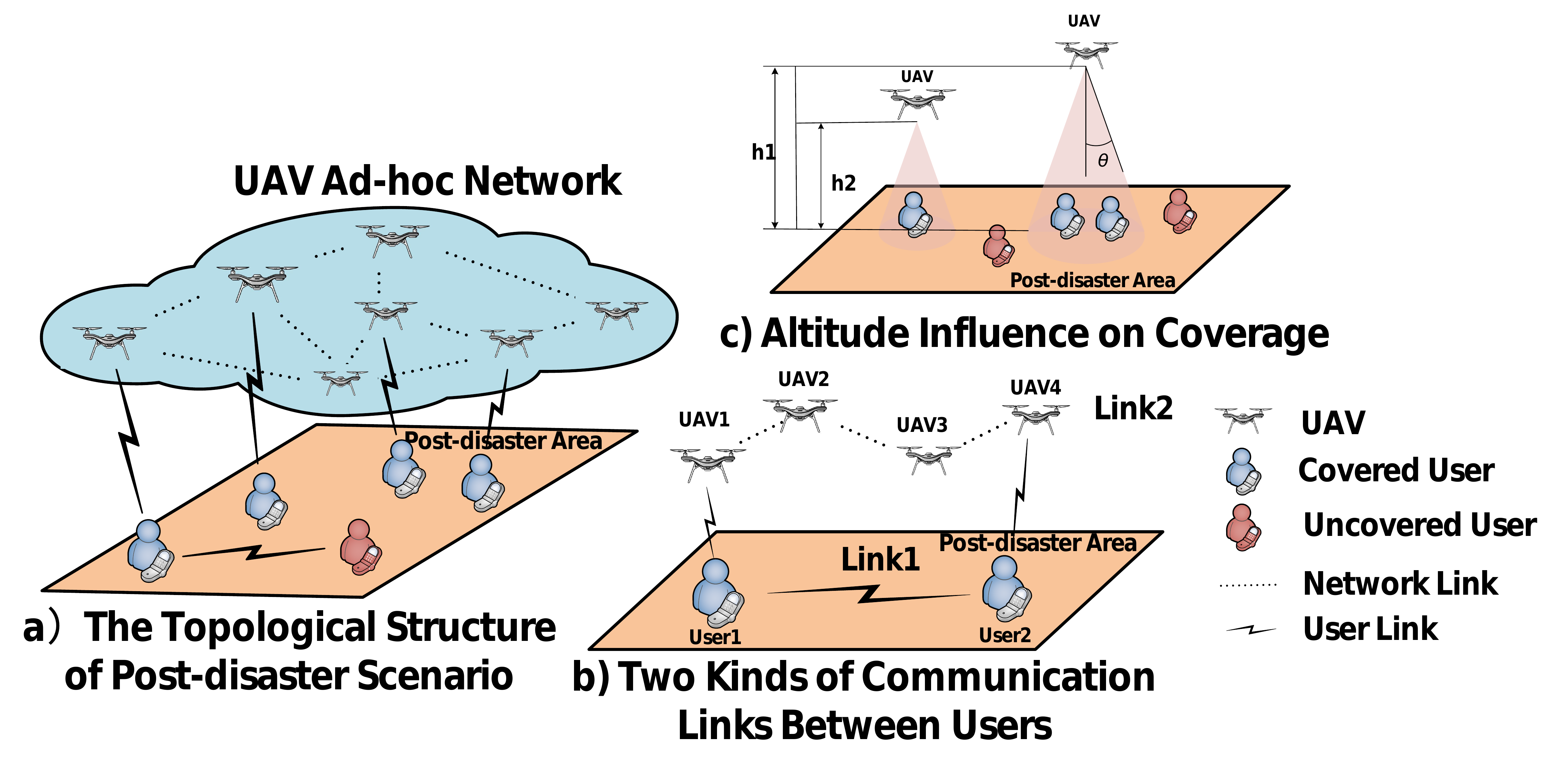}
\caption{The topological structure of UAV ad-hoc networks. a) The UAV ad-hoc network supports user communications. b) The coverage of a UAV depends on its altitude and field angle. c) There are two kinds of links between users, and the link supported by UAV is better. }
\label{fig:structure}
\end{figure}

\section{System model and Problem formulation}

\subsection{System Model}
We construct a UAV ad-hoc network in a post-disaster scenario with $M$ identical UAVs being randomly deployed, in which $M$ is a huge number compared with normal Multi-UAV system. All the UAVs have the same volume of battery $E$ and communication capability. The topological structure of Multi-UAV network is shown in Fig.~\ref{fig:structure} (a).

To support the communication mission, all UAVs are required to cooperate and support the user communication in need. UAVs work above post-disaster area $D$. If a user (${\rm User}_1$) needs to communicate with another user (${\rm User}_2$) far away from him/her, the communication link can be set either directly between ${\rm User}_2$ (${\rm User}_1\rightarrow{\rm User}_2$), or use UAVs above as relay (${\rm User}_1\rightarrow{\rm UAV}_1\rightarrow{\rm UAV}_2\rightarrow\cdots\rightarrow{\rm UAV}_n\rightarrow{\rm User}_2$), as shown in Fig. \ref{fig:structure} (b). Since the users are far away from each other, the link through UAVs provides better communication quality than the direct one, users will choose Link$2$ if they are covered by UAVs. Notice that users may not be covered by UAVs due to the high uncertainty in post-disaster areas.

\par In the UAV ad-hoc network, there are $N$ channels available for UAVs to communicate, which are labeled as $C=[C_1,...,C_n,...,C_N]$. Each channel holds at most $C_{max}$ UAVs due to interference constraint. Following the water-filling allocation scheme \cite{BinL}, UAVs select one channel $C_n$ in the beginning of the process and will not change their decision afterwards. Assuming that ${\rm UAV}_i$'s decision is represented by the vector $C_i=[C_{i1},...,C_{in},...C_{iN}]$, where $C_{in}=1$ if UAV$_i$ selects channel $C_n$, otherwise $C_{in}=0$. Since the channel of each UAV is fixed, UAVs cannot switch to other channel with less noise to improve SNR, but only capable of raising SNR by coordinating their own power. The power for ${\rm UAV}_i$ in each channel is denoted by the vector $P_i=[P_{i1},...,P_{in},...,P_{iN}]$, $P_{in}>0$ when ${\rm UAV}_i$ selects the channel $C_n$, otherwise, $P_{in}=0$. The intrinsic noise in each channel is $Noise=[Ns_1,...,Ns_n,...,Ns_N]$.

\par UAVs have several power levels and altitude levels. In the midst of extreme environments, UAVs cannot change its voltage dramatically but merely change to the adjacent power level \cite{TunG2}. Similarly, the altitude changing also has a limitation that only adjacent altitude level conversion is permitted in each move. We denote power set and altitude set to be $P=\{P_1,...,P_{k},...,P_{np}\}$ and $h=\{h_1,...,h_{k},...,h_{nh}\}$, respectively, where $np$ is the number of power levels, and $nh$ is the number of altitude levels. We assume that the gap between different levels of power and altitude are equal. Let $\Delta P$ and $\Delta h$ denote the distance value of adjacent power levels and altitude levels, respectively.

\par When UAVs need communications, and the signal to noise rate (SNR) mainly determines the quality of service. UAVs' power and inherent noise are interferences for each other. Since there are hundreds of UAVs in the system, each UAV is unable to sense all the other UAVs' power explicitly, but only sense and measure aggregative interference and treat it as an integral influence. Though increasing power can improve SNR, excessively large power causes more energy consumption and results in less running time. Therefore, proper power control for UAVs is needed to be carefully designed.

\par Coverage is another factor which determines the performance of each UAV. As presented in Fig.~\ref{fig:structure} (c), the altitude of UAV plays an important role in coverage adjusting. The higher altitude it is, the larger coverage size a UAV has. A large coverage size means a substantial opportunity of supporting more users, but a higher SNR will be needed. Furthermore, the turbulence of upper air disrupts the stability of UAVs with more energy consumption. Thus, a suitable height is essential to determine the coverage area.

\subsection{Problem Formulation}
Defining a UAV ad-hoc network game $ \Gamma=(U_i,S_i)_{i\in M} $, where $U_i$ is the utility function of ${\rm UAV}_i$ and $S_i$ is the strategy set of ${\rm UAV}_i$. Take $s_i\in S_i$ as one strategy of ${\rm UAV}_i$, then $s_i=[C_i,P_i,h_i]$. For simplicity, we denote $S_{-i}\in \prod _{j\neq i}S_j$ as the strategy set of UAVs except ${\rm UAV}_i$ and $s_{-i}\in S_{-i}$. $S=\prod _i S_i$, and $s\in S$ is the strategy profile of the game.

\par As is described previously, UAVs can only sense the aggregative influence in each channel they select. Defining such influence as a power interaction term, which of ${\rm UAV}_i$ is written as:
\begin{equation}
\sigma_{ip}(s_{-i})=(\sigma-P_i)\otimes C_i,
\end{equation}
where $\sigma=\sum_{i\in M}P_i+Noise$ and $\otimes$ is the operator that multiplies corresponding elements in two vectors. For the convenience of notation, denoting the $nth$ element in $\sigma_{ip}(s_{-i})$ to be $\sigma_{ip}(s_{-i})_{(n)}$, which is the power interaction in $C_n$. It should be noticed that, when ${\rm UAV}_i$ does not choose $C_n$, $\sigma_{ip}(s_{-i})_{(n)}=0$.

\par Supposing that a UAV covers a round area below it with a field angle $\theta$ as shown in Fig.~\ref{fig:structure} (b). Thus the coverage of ${\rm UAV}_i$ is $D_i=\pi(h_i{\rm tan}\theta)^2$. Considering that the higher the altitude is, the severer the air turbulence a UAV suffers, the utility of a coverage area is
\begin{equation}
\tilde{D}_i=\pi(h_i{\rm tan}\theta)^{2\cdot\beta},
\end{equation}
where $\beta$ is the air turbulence index, which decreases as the altitude increases.

\par
\begin{figure}[t]
\centering
\includegraphics[width=\linewidth]{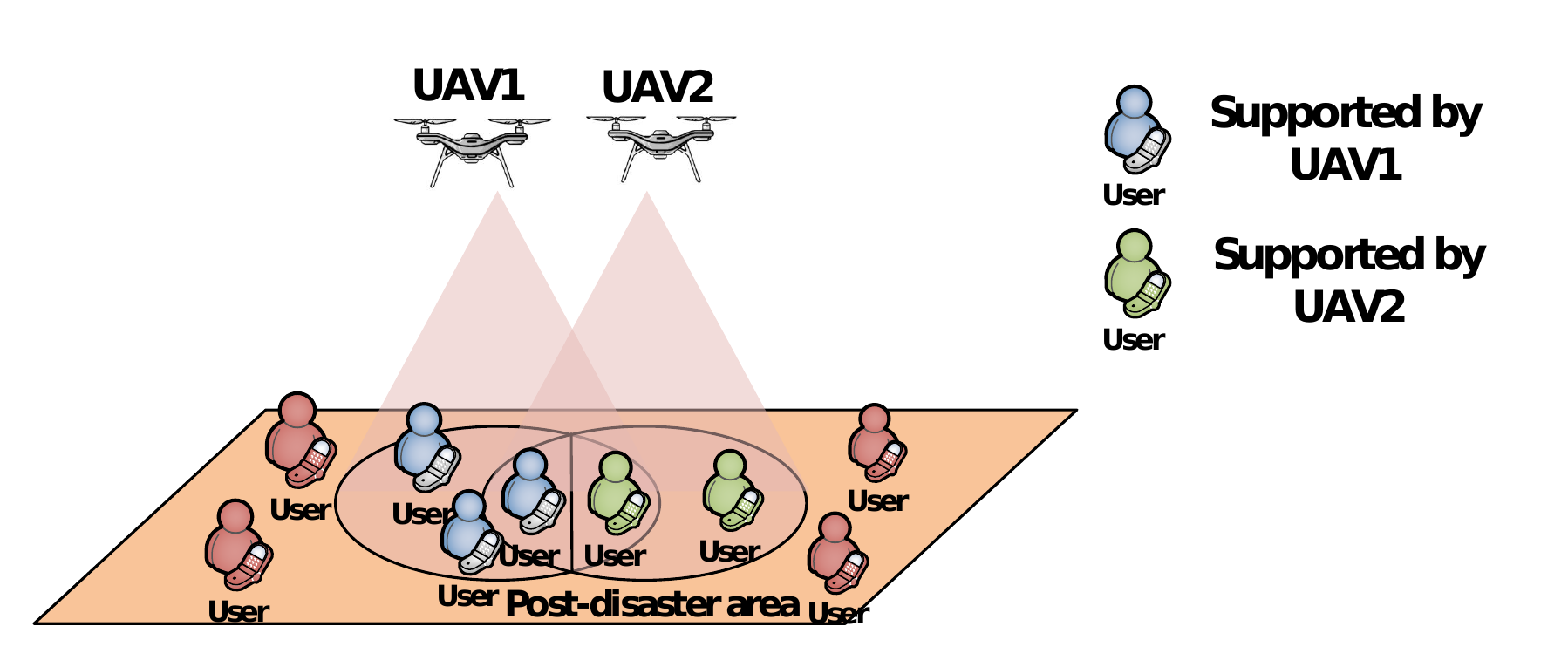}
\caption{Coverage overlap between two UAVs. When two UAVs are close, there will be overlap areas among them and the utility of coverage will decrease.}
\label{fig:overlap}
\end{figure}

 When there are numbers of UAVs in the network, it is possible for the coverage areas of different UAVs to overlap. When a UAV overlaps with another, they will not support all users but share the mission. The users in the overlaps will be served randomly with equal probability by each UAV. Fig.~\ref{fig:overlap} presents the overlaps between two UAVs, i.e. ${\rm UAV}_1$ and ${\rm UAV}_2$ will support half of the users in the overlap area. In this condition, the true coverage $\bar{D}_i$ is smaller than $D_i$, and it is written as
\begin{equation}
\bar{D}_i=D_i-\kappa\sum_{j\neq i}D_j,
\end{equation}
where $\kappa$ is the index which decides the influence of overlap. Since $\bar{D}_i$ must satisfy that $\bar{D}_i>0$, $\kappa$ is a tinny index. In a large-scale network, $ D_i<<\sum_{j\neq i}D_j$ and $\kappa<<1$.

\setlength{\abovecaptionskip}{0cm} 
\setlength{\belowcaptionskip}{-0.25cm} 

\par In order to support as many users as possible, UAVs are required to enlarge coverage size, which is equal to enlarge the coverage proportion in the mission area. Higher altitude indicates larger coverage size as shown in Fig.~\ref{fig:structure} (c). The utility of coverage size is denoted as
\begin{equation}
U_{D}=\frac{\tilde{D}_i}{D}.
\end{equation}

\par Now we define the utility functions of energy, SNR and coverage to calculate UAVs' payoffs.

\par For power selection of ${\rm UAV}_i$, a large power does not necessarily result in high utility due to the large interference comes with it. Taking energy saving and longer lifetime into consideration, choosing the right amount of power that balances the tradeoffs between power and interference, brings the highest utility. We define the energy utility function $U_{E}$ as:
\begin{equation}
U_{E}=\frac{E}{\sum\limits_{1\leq n\leq N}P_{in}}.
\end{equation}
It should be noted that a high power provides UAVs with higher communication quality and larger coverage requires higher power.

\par As for power control, the SNR in channel $C_n$ for ${\rm UAV}_i$ is
\begin{equation}
SNR=\frac{P_{in}}{\sum_{j\neq i}P_{jn}+Ns_n}.
\end{equation}
And it can also be written as $\mu C_{in}[P_{in}-\gamma\sigma_{ip}(s_{-i})_{(n)}]$ \cite{CheJ}, where $\gamma$ is the SNR balance index and $\mu$ is the SNR index. If given the Shannon capacity
\begin{equation}
C=Blog_2(1+SNR),
\end{equation}
which decides the communication quality to be improved, SNR is supposed to be enlarged. The summation SNR of ${\rm UAV}_i$ is $\mu\sum_{i\in M}[P_i-\gamma\sigma_{ip}(s_{-i})]\otimes C_i$, where we broaden the meaning of $\sum$ and the formula is the summation of all elements in the vector.

\par According to the previous definitions, we use an equation and an index to balance the tradeoff between power control and coverage, and we formulate the SNR utility function as:
\begin{equation}
U_{SNR}=\mu\sum_{i\in M}[P_i-\gamma\sigma_{ip}(s_{-i})]\otimes C_i-\alpha\bar{D}_i,
\end{equation}
where $\alpha$ is the index that balances the tradeoff between SNR and coverage size.

\par Finally, in terms of the utility function $U_i$ of ${\rm UAV}_i$, we should find a balance among power limitation, SNR and coverage size. Then the utility function $U_i$ of ${\rm UAV}_i$ is defined as:
\begin{footnotesize}
\begin{equation}
\begin{split}
&U_i(s_i,\sigma_{ip}(s_{-i}))=AU_{E}+BU_{SNR}+CU_{D}
\\&=\frac{AE}{\sum\limits_{1\leq n\leq N}P_{in}}+B\{\mu\sum[P_i-\gamma\sigma_{ip}(s_{-i})]\otimes C_i-\alpha\bar{D}_i\}+\frac{C\tilde{D}_i}{D},
\end{split}
\label{fun:utility}
\end{equation}
\end{footnotesize}
where $A$, $B$ and $C$ are balance indices that balance three utilities on the basis of post-disaster scenario. The ultimate goal for enlarging the utility of the networks is to enlarge the summation of utility function (\ref{fun:utility}) of each UAV, and we define the global utility function as the goal function:
\begin{equation}
U=\sum_{i\in M}U_i(s_i,\sigma_{ip}(s_{-i})),
\end{equation}
which presents the performance of the UAV ad-hoc network.

\section{Multi-aggregator Aggregative Game}
\subsection{Multi-aggregator Aggregative Game Model}
We formulate the UAV ad-hoc network game in large-scale post-disaster area as a multi-aggregator aggregative game \cite{JenM}, where we calibrate its definition in our UAV network model and put it as follows.
\begin{definition}(Multi-aggregator Aggregative Game) The game $\Gamma=(U_i,S_i)_{i\in M}$ is called a \emph{multi-aggregator aggregative game} with two or more aggregators $g={g_1,g_2,...,g_K}$, $g_k:S\rightarrow R^n$ ($n$ is the dimension of aggregators), and $S$ is bounded, if there exists two or more interaction functions $\sigma_i=\{\sigma_{i1},\sigma_{i2},...,\sigma_{iK}\}$, and $\sigma_{ik}: S_{-i}\rightarrow X_{-i}\subseteq R^n,i\in M$ such that each of the payoff functions $i\in M$ can be written as:
\begin{equation}
U_i(s)=u_i(s_i,\sigma_{i1}(s_{-i}),\sigma_{i2}(s_{-i}),...,\sigma_{iK}(s_{-i})),
\end{equation}
where $u_i:S_i\times X_{-i}\times X_{-i} \times ...\times X_{-i}\rightarrow R$, and for any fixed strategies profile $s$:
\begin{equation}
g_k(s)=g(\sigma_{ik}(s_i,s_{-i})).
\end{equation}
\end{definition}

\par In the UAV ad-hoc network game, $\sigma_{i1}(s_{-i})$ is the power interaction term $\sigma_{ip}(s_{-i})$, which is the aggregative power influence from other UAVs for ${\rm UAV}_i$. Another interaction term $\sigma_{i2}(s_{-i})$ is named as area interaction term $\sigma_{ia}(s_{-i})$:
\begin{equation}
\sigma_{i2}(s_{-i})=\sigma_{ia}(s_{-i})=\sum_{j\neq i}D_j.
\end{equation}

\par Besides, the aggregators $g_1(s)$ and $g_2(s)$ of the UAV ad-hoc network game are as follows:
\begin{equation}
g_1(s)_{(n)}=(\sigma_{ip}(s_{-i})+P_i)_{(n)},
\end{equation}
where $g_1(s)_{(n)}$ is the $n$th element of $g_1(s)$. $g_1(s)_{(n)}$ is the summation of power and noise in channel $n$, which has been chosen by ${\rm UAV}_i$. If the strategy profile $s$ is fixed, the summation of power and noise in all channels are fixed. Therefore, for any value of $i$, $g_1(s)_{(n)}$ is the same, if the stratefy profile $s$ is fixed. 
\begin{equation}
g_2(s)=\sigma_{ia}(s_{-i})+D_i.
\end{equation}
$\sigma_{ia}(s_{-i})$ is the area covered by UAVs except of ${\rm UAV}_i$. Therefore $g_1(s)$ is the summation of all UAV's coverage area. For any value of $i$, $g_2(s)$ is the same, if the stratefy profile $s$ is fixed. 

\par The utility function $U_i(S_i,\sigma_{ip}(S_{-i}))$ in UAV ad-hoc network game conforms to the definition of payoff function $U_i(S)$. Therefore, UAV ad-hoc network game is a multi-aggregator aggregative game.

\subsection{Analysis of Nash Equilibrium}
In game theory, Nash Equilibrium (NE) is a special state that no UAV can gain more payoff by changing its strategy. Thus, NE is an ideal solution for all UAVs in multi-UAV relay mission game \cite{CheJ}. The potential game is usually used in analyzing the existence of NE. We first define the Pure Strategy Nash Equilibrium (PSNE) with several assumptions, then prove that the UAV ad-hoc network game has a NE.
\begin{definition}(Pure Strategy Nash Equilibrium) A strategy profile $s^*=\{s_1^*,s_2^*,...,s_m^*\}$ is a \emph{pure strategy Nash equilibrium} if and only if any UAV is unable to gain more payoff by altering their strategies when no other UAVs change their strategies, i.e.,
\begin{equation}
U_i(s_i,s_{-i}^*)\leq U_i(s_i^*,s_{-i}^*), i\in M,s_i\neq s_i^*.
\end{equation}
\end{definition}

\begin{definition}(Potential Game) A game $\Gamma=(\phi_i,S_i)_{i\in M}$ such that $S_i$ is the strategy of ${\rm UAV}_i$, $s_i\in S_i$. We broaden the meaning of $\prod$ and write the set of all strategies as $S=\prod S_i$. Labeling the set of UAVs except ${\rm UAV}_i$ as $S_{-i}=\prod_{j\neq i}S_j$,$s_{-i}\in S_{-i}$. There exists a potential function $\phi_i(s_{-i}):S_{-i}\rightarrow S_i$. If there exists a function $P:S\rightarrow R$ for $\forall i\in M$ such that
\begin{equation}
U_i(s'_i,s_{-i})-U_i(s_i,s_{-i})=\phi(s'_i,s_{-i})-\phi(s_i,s_{-i}),
\end{equation}
then this game is a potential game.
\end{definition}
\par Definition 3 indicates that the change of utility function is the same amount of the change of potential function, which gives an ideal property to the potential game.

\begin{theorem}
Any potential game with finite strategy has at least one PSNE.
\end{theorem}
\IEEEproof
Please refer to the proof of Corollary 2.2 in \cite{potential}.
\endIEEEproof

\begin{theorem}
The UAV ad-hoc network game is a potential game and has at least one PSNE.
\end{theorem}
\IEEEproof
Please refer to Appendix A.
\endIEEEproof

\par Since the UAV ad-hoc network game is a special type of potential game, we can apply the properties of the potential game in the later analysis. Some algorithms that have been applied in the potential game can also be employed in the UAV ad-hoc network game. In the next section, we investigate the existing algorithm with its learning rate in large-scale post-disaster scenarios and propose a new algorithm which is more suitable for the UAV ad-hoc network in such scenarios.

\section{Learning Algorithm}
\subsection{Payoff-based Binary Log-Linear Learning}
In the literatures, most works search PSNE by using the Binary Log-linear Learning Algorithm (BLLA). However, there are limitations of this algorithm. In BLLA, each UAV can calculate and predict its utility for any $s_i\in S_i$ in the complete strategy set. While in UAV ad-hoc networks, UAVs are accessible only to the constrained strategy set and corresponding utilities in the last two decision periods. Thus, conventional BLLA is no more suitable for the scenario we considered here, we propose a revised algorithm based on BLLA, called Payoff-based Binary Log-linear Learning Algorithm (PBLLA) to resolve the issue.

\par In this section, we will present how PBLLA works and the convergence of PBLLA.

\begin{algorithm}
    \caption{Payoff-based Binary Log-linear Learning Algorithm}
    \begin{algorithmic}[1]
        \STATE \textbf{Initialization:} Selecting an arbitrary power and altitude profile $s\in S$ and arbitrary channels for each UAV, the number of UAVs in a channel must be less than $C_{max}$.
        \STATE Set $t=1$.
        \STATE Set $s(t)=s$, $x_i(t)=0$ for any ${\rm UAV}_i$ such that $i\in M$ .
        \STATE Set dynamic degree index $\tau\in(0,\infty)$.
        \STATE \textbf{Repeat}
        \IF{$x_i(t)=0$ for $\forall i\in M$}
            \STATE Select ${\rm UAV}_i$ randomly.
            \STATE Select $s_i(t+1)$ randomly in $C_i(s_i(t))$, which is the constrained strategies set when the current strategy is $s_i(t)$.
            \STATE $x_i(i+1)=1$.
        \ELSE
            \STATE Select ${\rm UAV}_i$ which satisfies $x_i(i+1)=1$,
            \STATE With probability $\frac{\exp[\frac{1}{\tau}U_i(s(t-1))]}{\exp[\frac{1}{\tau}U_i(s(t-1))]+\exp[\frac{1}{\tau}U_i(s(t))]}$,
            \STATE \quad$s_i(t+1)=s_i(t-1)$;
            \STATE Or with probability $\frac{\exp[\frac{1}{\tau}U_i(s(t))]}{\exp[\frac{1}{\tau}U_i(s(t-1))]+\exp[\frac{1}{\tau}U_i(s(t))]}$,
            \STATE \quad$s_i(t+1)=s_i(t)$.
            \STATE $x_i(t+1)=0$.
        \ENDIF
        \FOR {${\rm UAV}_j$, $j\in M$ and $j\neq i$}
        \STATE $s_j(t+1)=s_j(t)$ and $x_j(t+1)=0$
        \ENDFOR
        \STATE $t=t+1$
        \STATE \textbf{End Repeat}
    \end{algorithmic}
\end{algorithm}

\begin{theorem}
Let denote $\tau$ as the dynamic degree of the scenarios. The harsher environment the networks suffers, the higher $\tau$ it is. In the highly dynamic scenarios, we suppose that $\tau\geq0.01$. With proper $\tau$, PBLLA asymptotically converges and leads the UAV ad-hoc network game approaching to the PSNE.
\end{theorem}

\IEEEproof The proof of the convergence of PBLLA is similar to that of BLLA in \cite{SPBLLA}. According to \cite{SPBLLA}, we only need to illustrate that UAV ad-hoc network game conforms to the following two assumptions.

\begin{assumption}
For $\forall i\in M$ and all strategy profile pairs $s_i^0,s_i^n\in A$, there exists a series of strategies $s_i^0\rightarrow s_i^1 \rightarrow ... \rightarrow s_i^n$ satisfying that $s_i^k\in C_i(s_i^{k-1})$ for $\forall k\in \{ 1,2,...,n\}$, where $C_i(s_i^{k})$ is the constrained strategies set of strategy profile $s_i^{k}$.
\end{assumption}

\begin{assumption}
For $\forall i\in M$ and all strategy profile pairs $s_1^0,s_2^n\in A$,
\begin{equation}
s_i^2\in C_i(s_i^1)\Leftrightarrow s_i^1\in C_i(s_i^2).
\end{equation}
\end{assumption}

In the UAV ad-hoc network game, UAVs are only permitted to select the adjacent power and altitude levels. It is evident that for any strategy profile pair $s_i^0,s_i^n\in A$, $s_i^0$ can change power and altitude step by step to reach $s_i^n$, vice versa. Thus, UAV ad-hoc network game satisfies Assumption 1 and Assumption 2. The remaining proof can follow Theorem 5.1 in \cite{SPBLLA}.
\endIEEEproof

\par The essence of PBLLA is selecting an alternative UAV randomly in one iteration and improving its utility by altering power and altitude with a certain probability, which is determined by the utilities of two strategies and $\tau$. UAV prefers to select the power and altitude which provide higher utility. Nevertheless, highly dynamic scenarios will cause UAVs to make mistakes and pick the worse strategy. The dynamic degree index $\tau$ determines the dynamic degree of the situation and UAV's performance. Small $\tau$ means less dynamic scenarios and fewer mistakes when UAVs are making decisions. When $\tau\rightarrow0$ which equals to stabilization, UAV will always select the power and altitude with higher utility; when $\tau\rightarrow\infty$ where exists sever dynamics, UAV will choose them randomly. However, PBLLA has its limitations that PBLLA is only one single UAV is allowed for altering strategies in one iteration. We will propose a new algorithm in the next section to overcome the restrictions.

\subsection{Synchronous Payoff-based Binary Log-linear Learning}

Since PBLLA only allows one single UAV to alter strategies in one iteration, such defect would cause computation time to grow exponentially in large-scale UAVs systems. In terms of large-scale UAVs ad-hoc networks with a number of UAVs denoted as $M$, $M^2$ times of exchange messages will be needed to coordinate and guarantee that only one UAV changes strategy in each iteration. Such a process not only consumes large energy but also prolongs convergence time. Algorithms that can improve the learning rate and reduce messages exchange is urgently needed. Thus, we propose the Synchronous Payoff-based Binary Log-linear Learning Algorithm (SPBLLA), which permits each UAV altering their strategies synchronously and learning with no message exchange.

\begin{algorithm}
    \caption{Synchronous Payoff-based Binary Log-linear Learning Algorithm}
    \begin{algorithmic}[1]
        \STATE \textbf{Initialization:} Selecting an arbitrary power and altitude profile $s\in S$ and arbitrary channels for each UAV, the number of UAVs in a channel must be less than $C_{max}$.
        \STATE Set $t=1$$s(1)=s$
        \STATE Set $x_i(t)=0$ for any ${\rm UAV}_i$ such that $i\in M$.
        \STATE Set dynamic degree index $\tau\in(0,\infty)$.
        \STATE Set probability index $m$, and the altering strategies probability $\omega=(e^{-\frac{1}{\tau}})^m$.
        \STATE \textbf{Repeat}
        \FOR {$\forall {\rm UAV}_i$}
            \IF{$x_i(t)=0$ }
                \STATE With probability $\omega$:
                    \STATE \quad ${\rm UAV}_i$ select $s_i(t+1)$ randomly in $C_i(s_i(t))$, which is the constrained strategy set when the current strategy is $s_i(t)$,
                    \STATE \quad $x_i(i+1)=1$;
                \STATE or With probability $(1-\omega)$:
                    \STATE \quad $s_i(t+1)=s_i(t)$,
                    \STATE \quad $x_i(i+1)=0$.
            \ELSE
            \STATE With probability $\frac{\exp[\frac{1}{\tau}U_i(s(t-1))]}{\exp[\frac{1}{\tau}U_i(s(t-1))]+\exp[\frac{1}{\tau}U_i(s(t))]}$,
            \STATE \quad$s_i(t+1)=s_i(t-1)$;
            \STATE Or with probability $\frac{\exp[\frac{1}{\tau}U_i(s(t))]}{\exp[\frac{1}{\tau}U_i(s(t-1))]+\exp[\frac{1}{\tau}U_i(s(t))]}$,
            \STATE \quad$s_i(t+1)=s_i(t)$.
            \STATE $x_i(t+1)=0$.
        \ENDIF
    \ENDFOR
\STATE $t=t+1$.
\STATE \textbf{End Repeat}
\end{algorithmic}
\end{algorithm}

\par The process of SPBLLA let UAVs free from message exchange. Therefore, there is no waste of energy or time consumption between two iterations, which significantly improves learning efficiency. All UAVs are altering strategies with a certain probability of $\omega$, which is determined by $\tau$ and $m$. $\tau$ also presents the dynamic degree of scenarios. The chance of UAVs to make mistakes when altering strategies is determined by the dynamic degree as in PBLLA.
\par To prove the convergence of SPBLLA, we first provide some conceptions.

\begin{definition}(Regular Perturbed Markov Process)
Denote $P$ as the transaction matrix of a Markov Process which has a finite state space $S$. This Markov Process is called regular perturbed markov process with noise $\epsilon$ if the following conditions are met.
\\1) $P_\epsilon$ is aperiodic and irreducible when $\epsilon>0$.
\\2) $\lim_{\epsilon\rightarrow0}P_\epsilon=P_0$, where $P_0$ is an unperturbed process.
\\3) For any $s^n,s^m\in S$, when $P_\epsilon(s^n,s^m)>0$, there exists a function $R(s^n\rightarrow s^m)\geq0$, which is called the resistance of changing strategy from $s^n$ to $s^m$, such that
\begin{equation}
0<\lim_{\epsilon\rightarrow0+}\frac{P_\epsilon(s^n,s^m)}{\epsilon^{R(s^n\rightarrow s^m)}}<\infty.
\end{equation}
\end{definition}

\begin{definition}(Stochastically Stable Strategy)
Denote $P_\epsilon$ as the transaction probability of a regular perturbed Markov process in a state space $S$, and $\mu_\epsilon(s)$ is the probability that the state transforms to $s$. The state is a stochastically stable strategy if
\begin{equation}
\lim_{\epsilon\rightarrow0+}\mu_\epsilon(s)>0.
\end{equation}
\end{definition}

\par In our model, when UAVs of repeated UAV ad-hoc network game adheres to regular perturbed Markov process, the probability of being in $s$ is
\begin{equation}
\mu_\epsilon(s)=\frac{\epsilon^{-\phi(s)}}{\sum_{\tilde{s}\in S}\epsilon^{-\phi(\tilde{s})}}.
\end{equation}
Let $L$ denote a series of adjacent strategies, then $L=\{s^0\rightarrow s^1\rightarrow...\rightarrow a^n\}$. In any adjacent strategies pairs $(s^{k-1},s^k)$, the resistance of changing strategy from $s^{k-1}$ to $s^k$ is $R(s^{k-1},s^k)$. The resistance of path $L$ is the sum of each move
\begin{equation}
R(L)=\sum_{k=1}^{m}R(s^{k-1}\rightarrow s^k).
\end{equation}
According to $L$, we give the definition of resistance tree.

\begin{definition}(Resistance Tree)
In a strategy profile space $S$, strategy profiles are linked. A tree $T$, rooted at strategy profile $s$, is a set of directed edges that any other strategy profile has only one directed path, which consists of several directed edges, leads to $s$. The resistance of $T$ is the summation of resistance of all directed edges,
\begin{equation}
R(T)=\sum_{{s^{'}\rightarrow s^{''}}\in T}R(s^{'}\rightarrow s^{''}).
\end{equation}
Denoting $T(s)$ as all the trees that rooted at strategy profile $s$, then the stochastic potential of strategy profile $s$ can be written as
\begin{equation}
\gamma(s)=\min_{T\in T(s)}R(T).
\end{equation}
The minimum resistance tree of strategy space $S$ is the tree that has the minimum stochastic potential,
\begin{equation}
R(T_{min})=\min_{s\in S}\gamma(s).
\end{equation}
\end{definition}
\par
\begin{figure}[t]
    \centering  
    \includegraphics[width=\linewidth]{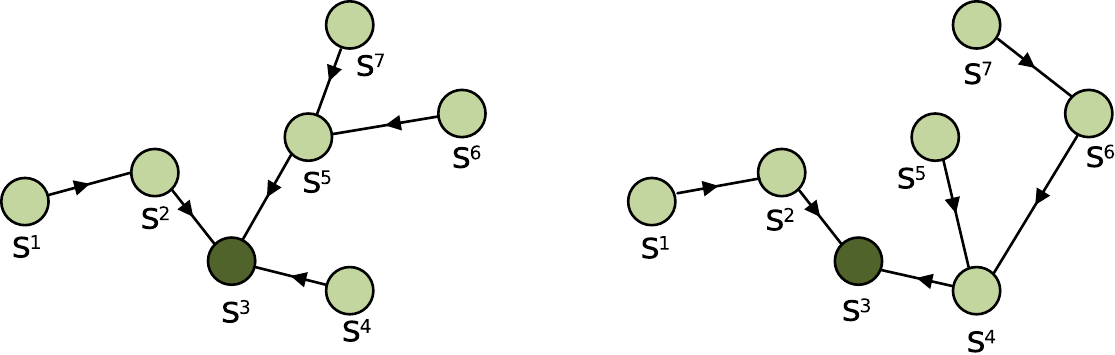}  
    \caption{Two trees rooted at $s_3$}  
    \label{fig:tree}   
\end{figure}

 For example, Fig.~\ref{fig:tree} shows different tree rooted at $S^3$. The branches are linked in different ways.

\setlength{\abovecaptionskip}{0cm} 
\setlength{\belowcaptionskip}{-0.25cm} 

\par With these definitions, we can prove the convergence of SPBLLA.

\begin{theorem}
Considering several UAVs with UAV ad-hoc network game with potential function $\phi:S\rightarrow R$. When all UAVs adhere to SPBLLA, if $m$ is large enough, the stochastically stable strategies are maximizers of the potential function, which are PSNEs.
\end{theorem}

\IEEEproof
Refer to Appendix B.
\endIEEEproof
\begin{remark}
According to Appendix B, to make the SPBLLA converge, $m$ should be twice larger than the most massive altering amount of each UAV's utility function.
\end{remark}

\begin{theorem}
$m$ is an index that indirectly influences the learning rate. If $m$ satisfies
\begin{equation}
m>2\Delta,
\label{m}
\end{equation}
where
\begin{small}
\begin{equation}\nonumber
\begin{split}
\Delta=&AE\frac{\Delta P}{\sum C_i\cdot P_1\cdot (P_1+\Delta P)}+B\sum C_i\cdot\Delta P
\\&+B\gamma\Delta P\cdot\sum(\sum\limits_{j\neq i}C_j\otimes C_i)
\\&+B\alpha\pi tan^2\theta(2h_{nh}\Delta h-\Delta h^2)
\\&+B\alpha\kappa(M-1)\pi tan^2\theta(2h_{nh}\Delta h-\Delta h^2)
\\&+\frac{C}{D}\pi tan^2\theta[h_{nh}^{2\beta}-(h_{nh}-\Delta h)^{2\beta}],
\end{split}
\end{equation}
\end{small}
then SPBLLA converge.
\label{mmax}
\end{theorem}
\IEEEproof
Refer to Appendix C.
\endIEEEproof

\begin{remark}
Let each UAV alter strategy as large as possible to make utility function change significantly. Calculating the most significant difference that a utility function can make in an iteration, and we are capable of learning the range of $m$.
\end{remark}

\setlength{\abovecaptionskip}{0cm} 
\setlength{\belowcaptionskip}{-1cm} 
\begin{figure*}
\begin{minipage}[t]{0.32\linewidth}
\centering
\includegraphics[width=\linewidth]{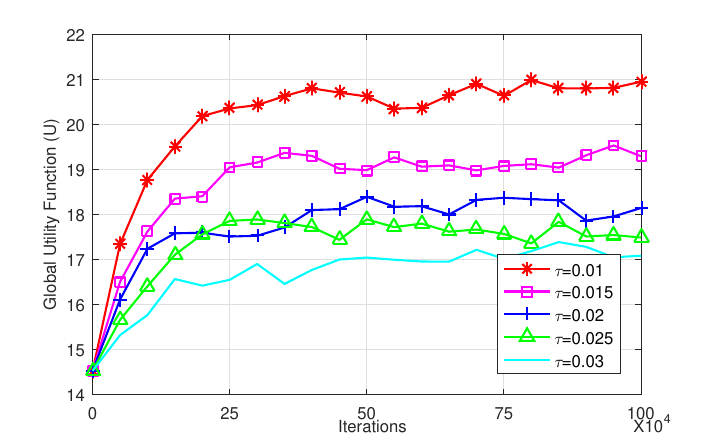}
\caption{Effect of dynamic degree index $\tau$ on PBLLA ($10^6$ iterations).  Severe dynamic scenarios cause fewer utilities of the whole networks.}
\label{fig:tauPBLLA1}
\end{minipage}%
\hspace{0.1cm}
\begin{minipage}[t]{0.32\linewidth}
\centering
\includegraphics[width=\linewidth]{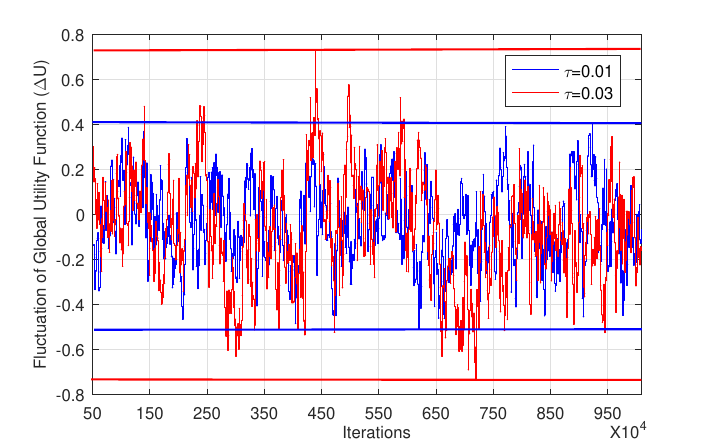}
\caption{Effect of dynamic degree index $\tau$ on PBLLA's fluctuation. Two horizontal lines show the fluctuation range. Higher dynamic scenario causes severe fluctuation.}
\label{fig:tauPBLLA2}
\end{minipage}
\hspace{0.1cm}
\begin{minipage}[t]{0.32\linewidth}
\centering
\includegraphics[width=\linewidth]{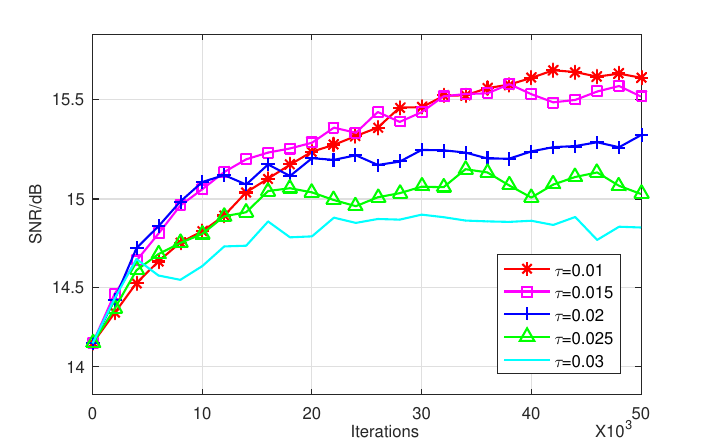}
\caption{Performance of SNR. In the ower dynamic scenario, SNR is much higher after convergence.}
\label{fig:SNR}
\end{minipage}
\end{figure*}

\setlength{\abovecaptionskip}{0cm} 
\setlength{\belowcaptionskip}{-0.5cm} 

\begin{figure*}
\begin{minipage}[t]{0.32\linewidth}
\centering
\includegraphics[width=\linewidth]{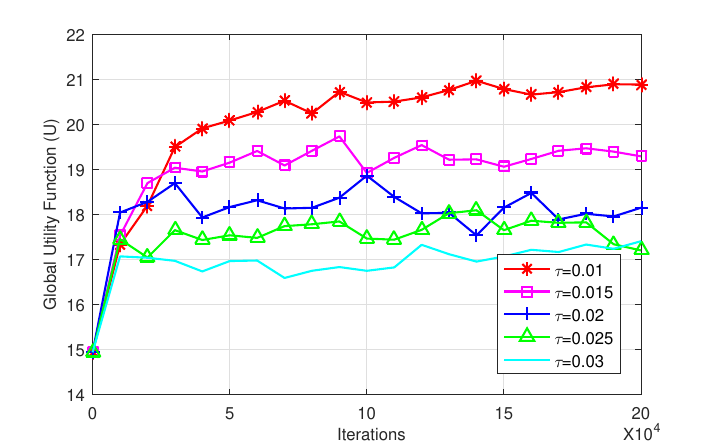}
\caption{Effect of dynamic degree index $\tau$ on SPBLLA ($2\times10^5$ iterations). The result is the same as PBLLA, which illustrates that algorithm does not affect convergence states.}
\label{fig:tauSPBLLA1}
\end{minipage}%
\hspace{0.1cm}
\begin{minipage}[t]{0.32\linewidth}
\centering
\includegraphics[width=\linewidth]{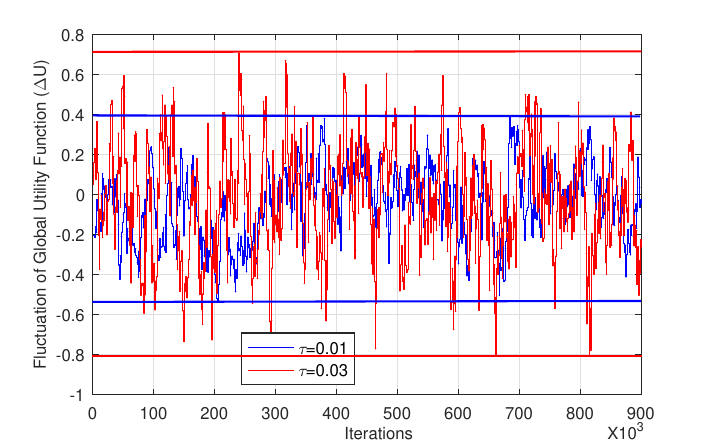}
\caption{Effect of dynamic degree index $\tau$ on SPBLLA ($2\times10^5$ iterations). The result is the same as PBLLA, which illustrates that algorithm does not affect convergence states.}
\label{fig:tauSPBLLA2}
\end{minipage}
\hspace{0.1cm}
\begin{minipage}[t]{0.32\linewidth}
\centering
\includegraphics[width=\linewidth]{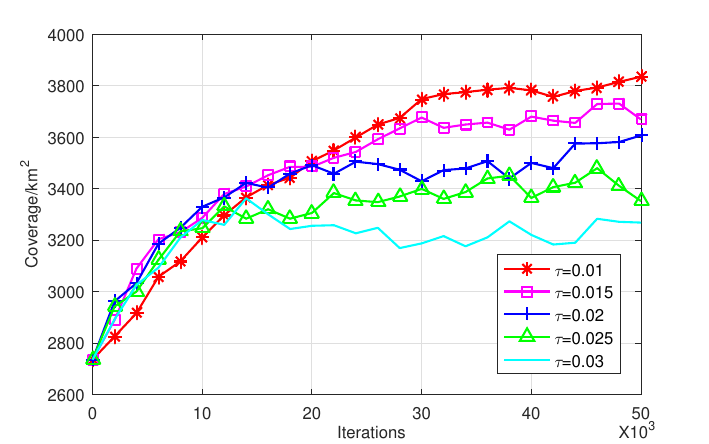}
\caption{Performance of Coverage size. Lower dynamic degree index increases Coverage size. When $\tau=0.01$, coverage proportion is up to 95\%.}
\label{fig:COVER}
\end{minipage}
\end{figure*}

\section{Simulation Result and Discussion}
In this section, how the key parameters in the UAV ad-hoc network affect the performance of PBLLA and SPBLLA will be studied. In the simulation, besides the quantity of UAVs and channel, other parameters are fixed as constant values. We set up the post-disaster area as $D=4000km^2$. $M=100$ UAVs are scatted in the area. There are $N=30$ channels and each channel can hold $C_{max}=25$ UAVs at most, and each UAV selects $N_C=5$ channels set $np=40$ levels of power are accessible $P=\{0.025W,0.05W,0.75W,...,0.975W,1W\}$, and the gap between two levels is $\Delta P=0.025W$. Noise is randomly distributed from $0.025W$ to $1W$. $nh=46$ levels of altitude are accessible $h=\{1km,1.2km,1.4km,...,9.8km,10km\}$, and the distinction between two levels is $\Delta h=0.2km$. The battery capacity of UAV for communication is $E=5mAh$. Field angle which determines the coverage size is $\theta=30^o$. The balance indices vary from one to another in the diverse scenarios. Here we choose $A=0.002$, $B=0.005$, $C=0.03$, and $\alpha=0.002$, $\gamma=0.002$, $\kappa=10^{-4}$, $\mu=10$. $\beta$ decreases as $h$ increases. According to Theorem \ref{mmax} in the last section, when $m>2\Delta$, SPBLLA is feasible.

\subsection{Effect of Scenarios' Dynamic Degree $\tau$}

In this part, we investigate the influence of environment dynamic on the network states. With different scenarios' dynamic degree $\tau\in(0,\infty)$, PBLLA and SPBLLA will converge to the maximizer of goal function with different altering strategy probability. Fig.~\ref{fig:tauPBLLA1} presents the influence of the dynamics on PBLLA. We can find out that the fluctuation during converging is severe in both algorithms, which is different from other related works. It does not result from the bad performance of algorithms but from the highly dynamic scenarios. When the environment is highly dynamic with high values of $\tau$, which brings about more mistakes when selecting powers and altitudes. Thus, when UAVs have low probabilities to select the right strategy, it will result in non-optimal decisions. In the rest simulations, similar phenomena can also be observed.

\par As the dynamic degree index $\tau$ decreases from $0.03$ to $0.01$, the goal function's values are increasing, which illustrates that lower values of $\tau$ approach to maximizer of the global utility function. When $\tau=0.03$, the value of $U$ does not increase much before converge. The fact that the severe interference from the environment seriously influences the UAV network making UAVs, making mistakes when altering strategies, can account for such a result. In Fig.~\ref{fig:tauPBLLA2}, $\Delta U$ is the value that presents the difference between the value of each iteration and the average function value after convergence. Compare two lines of $\tau=0.01$ and $\tau=0.03$ in Fig.~\ref{fig:tauPBLLA2}. The fluctuation of $\tau=0.03$ is around 50\% more than that of $\tau=0.01$, showing that the rough environments lead to unstable converge states. Fig.~\ref{fig:tauSPBLLA1} and~\ref{fig:tauSPBLLA2} show the performance of SPBLLA, which is similar to that of PBLLA. In the simulation of the SPBLLA, $m=0.03$, which follows the instruction of Theorem \ref{mmax}. Those mentioned figures show that when $\tau$s in PBLLA and SPBLLA are equal, the final optimum states and maximizer are similar.

\subsection{Performance on SNR and Coverage}
Since PBLLA is similar to SPBLLA, we skip the performance analysis for PBLLA, but focus on SPBLLA only. We set $m=0.03$ and $\tau$ from $0.01$ to $0.03$. Fig.~\ref{fig:SNR} and Fig.~\ref{fig:COVER} present the performance average SNR and the coverage size of the whole UAV ad-hoc network. The interference from the environment causes fluctuation. When the scenario dynamic degree is getting lower, the network is capable of approaching to a better outcome, and the SNR and coverage size is getting larger. In the slower dynamic scenario ($\tau=0.01$), the UAV ad-hoc networks can cover 95\% post-disaster area. Even though the interference from the environment is serious ($\tau=0.03$), the coverage proportion can be up to 80\%.

\subsection{Altering strategies probability $m$ on SPBLLA}
The opportunities for UAVs changing strategies in each SPBLLA's iteration are determined by the probability index $ m$. The altering probability $\omega=(e^{-\frac{1}{\tau}})^m$ illustrates that higher $m$ provide less probability for UAVs altering strategies. Higher probability enables higher chances for more UAVs altering strategies at the same time. Thus lower $m$ ensures high frequencies of changing strategies and faster converge rates. According to the limitation of $m$ in equ.~(\ref{m}), we figure out that $m>0.028$.

\par Fig.~\ref{fig:SPBLLAm} shows the effect of $m$ on the behavior of SPBLLA. Setting $\tau=0.01$ and $m>0.028$, we choose $5$ values from $0.03$ to $0.05$. As $m$ getting higher, SPBLLA needs more time for convergence. Since higher $m$ results in less opportunity for UAVs to alter strategies, fewer UAVs change strategies at the same time. For the sake of getting to the same optimum state, higher $m$ requires a longer time for convergence. It is also presented in Fig.~\ref{fig:SPBLLAm} that $m$ does not influence the optimum state. The same value of $\tau$ and various values of $m$ converge to the state that shares the similar value of the global utility function, and a higher probability for altering strategies save UAV ad-hoc networks convergence time.

\subsection{Power Control}
Allocating appropriate power is one of the key factors that affect UAV's utility, and the proper value of power will significantly improve the quality of service.

\par
\begin{figure*}
\begin{minipage}[t]{0.32\linewidth}
\centering
\includegraphics[width=\linewidth]{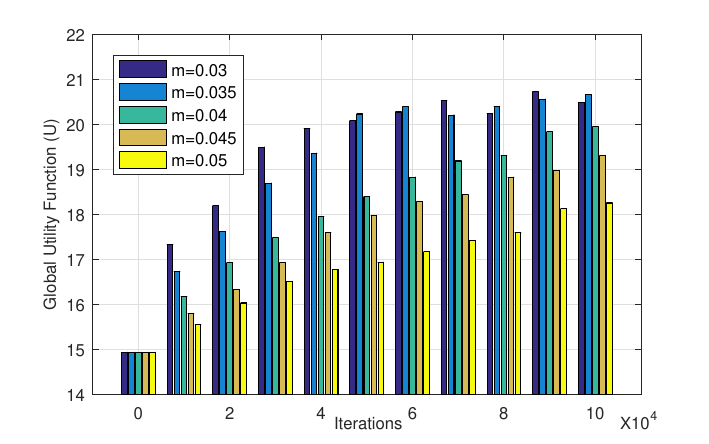}
\caption{Effect of probability index $m$ on SPBLLA. Lower probability indices create higher altering chances and higher learning rates. }
\label{fig:SPBLLAm}
\end{minipage}%
\hspace{0.1cm}
\begin{minipage}[t]{0.32\linewidth}
\centering
\includegraphics[width=\linewidth]{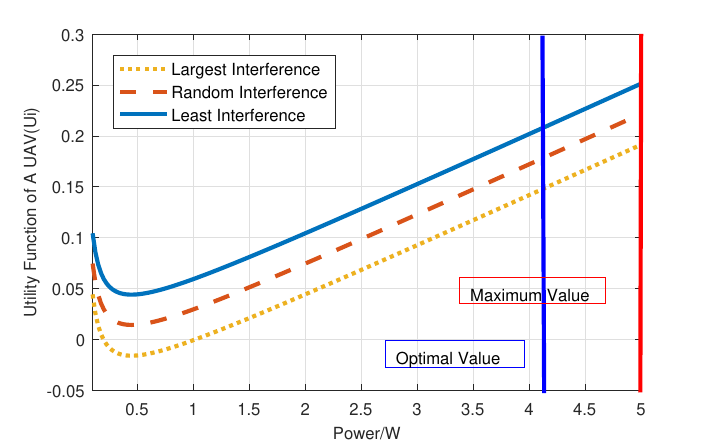}
\caption{Effect of power on the utility. The best power will locate in some values that smaller than the largest power.}
\label{fig:powerutility}
\end{minipage}
\hspace{0.1cm}
\begin{minipage}[t]{0.32\linewidth}
\centering
\includegraphics[width=\linewidth]{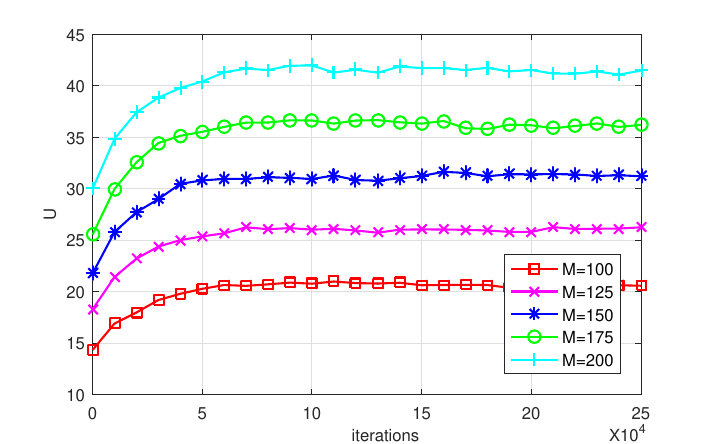}
\caption{Effect of UAVs quantity in SPBLLA. More UAVs provide more UAV ad-hoc networks utilities.}
\label{fig:numberSPBLLA}
\end{minipage}
\end{figure*}

 Fig.~\ref{fig:powerutility} presents the sketch diagram of a UAV's utility with power altering. The altitudes of UAVs are fixed. When other UAVs' power profiles are altering, the interference increases and the curve moves down. The high interference will reduce the utility of the UAV. Fig.~\ref{fig:powerutility} also shows that utility decreases and increases with power improving. Small and large power both provide high utilities, which is because small power will save energy and large power will increase SNR. The UAV might select the largest power to increase utility. However, The more power one UAV uses, the more interference other UAVs will receive and other UAVs' utilities will reduce. For the sake of enlarging the global utility, the largest power is not the optimal strategies for the whole UAV ad-hoc network. The best power will locate in some values that smaller than the largest power (The optimal value in the figure is a sketch value).


\setlength{\abovecaptionskip}{0cm} 
\setlength{\belowcaptionskip}{-0.5cm} 

\setlength{\abovecaptionskip}{0cm} 
\setlength{\belowcaptionskip}{-0.5cm} 

\begin{figure*}
\begin{minipage}[t]{0.32\linewidth}
\centering
\includegraphics[width=\linewidth]{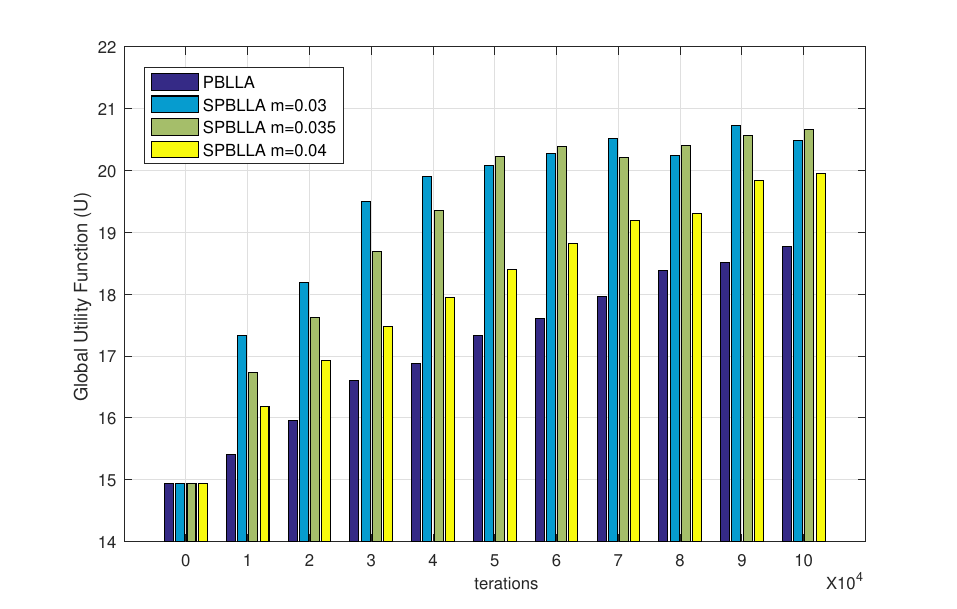}
\caption{Comparison of PBLLA with SPBLLA when $\tau=0.01$. With tree altering probabilities, SPBLLA's learning rates are much higher than that of PBLLA.}
\label{fig:compare1}
\end{minipage}%
\hspace{0.1cm}
\begin{minipage}[t]{0.32\linewidth}
\centering
\includegraphics[width=\linewidth]{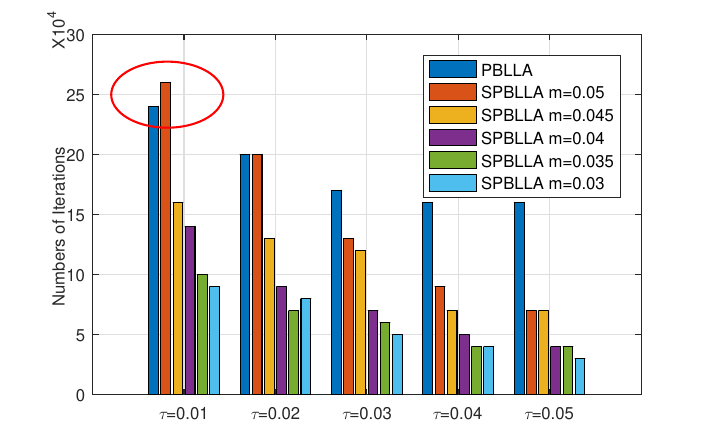}
\caption{Comparison of PBLLA with SPBLLA with various $\tau$. SPBLLA's learning rates are not always higher than those of PBLLA. The red circle shows the situation that SPBLLA is lower than PBLLA.}
\label{fig:compare2}
\end{minipage}
\hspace{0.1cm}
\begin{minipage}[t]{0.32\linewidth}
\centering
\includegraphics[width=\linewidth]{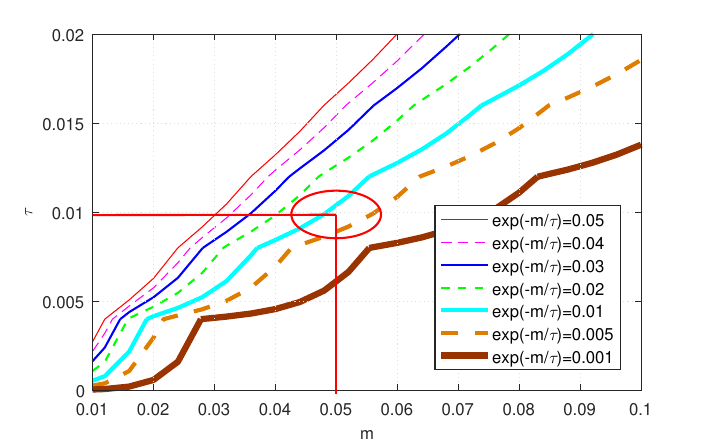}
\caption{$\tau$ and $m$'s impact on the probability of altering strategies $\omega=(e^{\frac{1}{\tau}})^m$. These lines are equal probability lines with various $\tau$ and $m$.}
\label{fig:tm}
\end{minipage}
\end{figure*}

\subsection{The number $N$ of UAVs}
In the large-scale UAV ad-hoc networks, the number of UAVs is another feature that should be investigated. Since the demanding channel's capacity should not be more than the channel's size we provide, we limit the number of UAVs in the tolerance range which satisfies that each UAV's channel selection is contented. In this scenario, there are $N=50$  channels, and the number of UAVs should be limited to $M\leq250$.

\par Fig.~\ref{fig:numberSPBLLA} shows how the number of UAVs affect the computation complexity of SPBLLA. Since the total number of UAVs is diverse, the goal functions are different. The goal functions' value in the optimum states increase with the growth in UAVs' number. Since goal functions are the summation function of utility functions,  more UAVs offer more utilities which result in higher potential function value. Moreover, more UAVs can cover more area and support more users, which also corresponds with more utilities. Fig.~\ref{fig:numberSPBLLA} also shows how many iterations that UAV ad-hoc network needs to approach to convergence. With the number of UAVs improves, more iterations are required in this network.


\subsection{Comparison of PBLLA and SPBLLA}
In this subsection, we do a comparison between PBLLA and SPBLLA to investigate the superiority of SPBLLA. We fix $\tau$ at several different values then compare the convergence rates of PBLLA and SPBLLA.

\par Fig.~\ref{fig:compare1} presents the learning rate of PBLLA and SPBLLA when $\tau=0.01$. As $m$ increases the learning rate of SPBLLA decreases, which has been shown in Fig.~\ref{fig:compare1}. However, when $m$ is small, SPBLLA's learning rate is about 3 times that of PBLLA showing the great advantage of synchronous learning. When $\tau=0.015$ and $\tau=0.02$ as shown in Fig.~\ref{fig:compare2}, such phenomenon also exists. Since PBLLA merely permits a single UAV to alter strategies in one iteration, SPBLLA's synchronous learning rate will much larger than PBLLA. Moreover, in the large-scale UAV network with high dynamic, PBLLA needs information exchange to decide the update order, which would severely prolong the learning time. PBLLA's learning time might be four times as long as that of SPBLLA. Thus we can make the conclusion that in the same condition (the same $\tau$ and other indexes), SPBLLA performs better and is more suitable for large-scale highly dynamic environment than PBLLA, and SPBLLA can improve the learning rate several times. With larger altering strategies probability, SPBLLA will be even more powerful.

\par However, we have to recognize that the altering strategies probability $\omega$ severely impacts on the efficiency of SPBLLA. If Theorem \ref{mmax} limits $m$ to be a large value, the probability will decrease. When $m$ is too large, UAVs are hard to move, and the learning rate will decrease. To some points, the learning rate of SPBLLA will lower than that of PBLLA. In our UAV ad-hoc network scenario, when $\tau=0.01$ and $m=0.03$, which is circled in Fig.~\ref{fig:compare2}, the probability of altering strategies $\omega<0.01$. The probability of altering strategies in SPBLLA is less than that of PBLLA, and the SPBLLA will spend more learning time.

\par Fig.~\ref{fig:tm} shows $\tau$ and $m$'s impact on the probability of altering strategies $\omega=(e^{\frac{1}{\tau}})^m$. The red circle in Fig.~\ref{fig:tm} matches with the red circle in Fig.~\ref{fig:compare2}, where SPBLLA is not efficient. The line $\omega=(e^{\frac{1}{\tau}})^m=0.01$ marks the same efficiency of the two algorithms, where SPBLLA alters one UAV in one iteration on average. When post-disaster scenarios are in the same degree of dynamic, lower $m$ permits higher changing strategies probability. If we want to increase altering probability, we can limit the utility change in each iteration to reduce $m$. Nevertheless, such a method reduces the updated amount which also causes a negative influence on the learning rate. Then how to balance the altering probability and update amount is a new topic that needs further investigation. Besides, how to reduce $m$ in Theorem \ref{mmax} is another task that requires further research.


\section{Conclusion}
In this paper, we establish a UAV ad-hoc network in the large-scale post-disaster area with the aggregative game. We propose a synchronous learning algorithm (SPBLLA) which expedites the learning rate comparing to the asynchronous learning algorithm (PBLLA), and shows desired behavior in highly dynamic scenarios. The learning rate of SPBLLA can be $10$ times larger than that of PBLLA. Even though there exists fluctuation during convergence, SNR has been improved and the network can cover over 95\% post-disaster area. From our analysis, we see that more UAVs provide higher utilities in the network. Though our proposed algorithm fits a highly dynamic environment, the learning rate of SPBLLA decreases when $m$ becomes large. In further work, improved SPBLLA which supports a wide range of $m$ will be needed.

\appendices
\section{Proof of the Theorem 2}
\IEEEproof
Formulate a function $\phi$:
\begin{small}
\begin{equation}
\phi(s)=\sum_{i\in M}\{A\frac{E}{\sum\limits_{1\leq n\leq N}P_{in}}+B[\sum\limits_{1\leq n\leq N}P_{in}-\alpha\bar{D}_i]+C\frac{\tilde{D}_i}{D}\}.
\end{equation}
\end{small}
\par When $UAV_i$ change its strategy from $s$ to $s'$, its utility changes
\begin{footnotesize}
\begin{equation}
\begin{split}
&U_i(s'_i,s_{-i})-U_i(s_i,s_{-i})
\\&=\Bigg(A\frac{E}{\sum\limits_{1\leq n\leq N}P'_{in}}+B\{\sum[P'_i-\gamma\sigma_{ip}(s_{-i})]\otimes C_i-\alpha\bar{D}'_i\}+C\frac{\tilde{D}'_i}{D}\Bigg)
\\&\quad-\Bigg(A\frac{E}{\sum\limits_{1\leq n\leq N}P_{in}}+B\{\sum[P_i-\gamma\sigma_{ip}(s_{-i})]\otimes C_i-\alpha\bar{D}_i\}+C\frac{\tilde{D}_i}{D}\Bigg)
\\&=A(\frac{E}{\sum\limits_{1\leq n\leq N}P'_{in}}-\frac{E}{\sum\limits_{1\leq n\leq N}P_{in}})+B(\sum(P'_i-P_i)\otimes C_i)
\\&
\quad+B\alpha[(D'_i-\kappa\sum_{j\neq i}D_j)-(D_i-\kappa\sum_{j\neq i}D_j)]+C(\frac{\tilde{D}'_i}{D}-\frac{\tilde{D}_i}{D})
\\&=A(\frac{E}{\sum\limits_{1\leq n\leq N}P'_{in}}-\frac{E}{\sum\limits_{1\leq n\leq N}P_{in}})+B\sum\limits_{1\leq n\leq N}(P'_{in}-P_{in})
\\&
\quad+B\alpha(D'_i-D_i)+C(\frac{\tilde{D}'_i}{D}-\frac{\tilde{D}_i}{D}).
\end{split}
\end{equation}
\end{footnotesize}
As for function $\phi$
\begin{small}
\begin{equation}
\begin{split}
&\phi(s'_i,s_{-i})-\phi(s_i,s_{-i})
\\&=\sum_{i\in M}\{A\frac{E}{\sum\limits_{1\leq n\leq N}P'_{in}}+B[\sum\limits_{1\leq n\leq N}P'_{in}-\alpha\bar{D}'_i]+C\frac{\tilde{D}'_i}{D}\}
\\&
\quad-\sum_{i\in M}\{A\frac{E}{\sum\limits_{1\leq n\leq N}P_{in}}+B[\sum\limits_{1\leq n\leq N}P_{in}-\alpha\bar{D}_i]+C\frac{\tilde{D}_i}{D}\}
\\&=A(\frac{E}{\sum\limits_{1\leq n\leq N}P'_{in}}-\frac{E}{\sum\limits_{1\leq n\leq N}P_{in}})+B\sum\limits_{1\leq n\leq N}(P'_{in}-P_{in})
\\&
\quad+B\alpha(D'_i-D_i)+C(\frac{\tilde{D}'_i}{D}-\frac{\tilde{D}_i}{D}).
\end{split}
\end{equation}
\end{small}
\par Therefore, $$U_i(s'_i,s_{-i})-U_i(s_i,s_{-i})=\phi(s'_i,s_{-i})-\phi(s_i,s_{-i})$$ and $\phi$ is the potential function of UAV ad-hoc network game.
\endIEEEproof

\begin{figure*}
    \centering  
    \includegraphics[width=\linewidth]{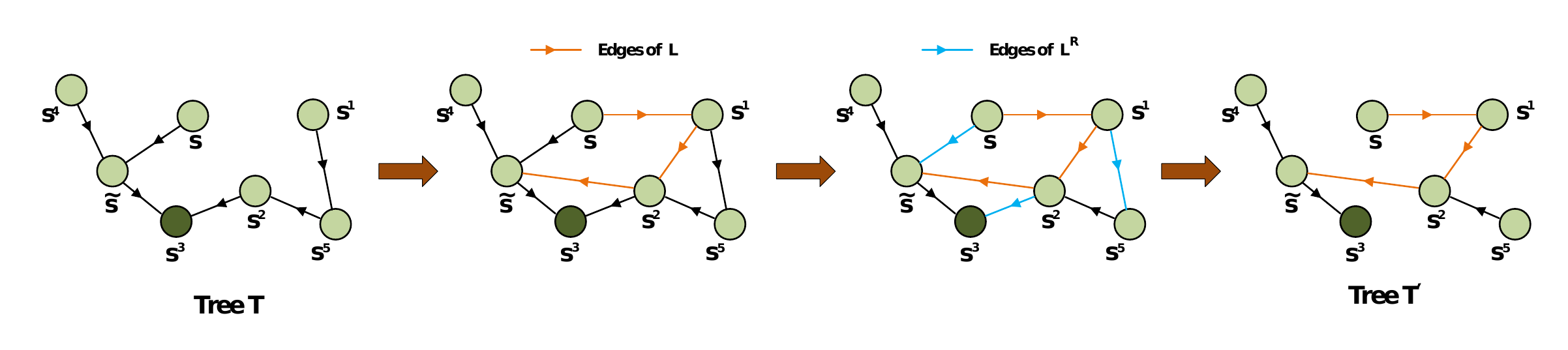}  
    \caption{Tree T is a tree that has an edge $s\rightarrow \tilde{s}$, where multiple UAVs atler strategies. We give a method of transformation from $T$ to $T'$, which removes $s\rightarrow \tilde{s}$: add orange edges of $L$ (edges $s\rightarrow s^1$, $s^1\rightarrow s^2$, and $s^2\rightarrow \tilde{s}$) to the tree T and then remove blue edges of $L^R$ (edges $s^1\rightarrow s^5$, $s^2\rightarrow s^3$, and $s\rightarrow \tilde{s}$) from the tree T.}  
    \label{fig:treeT'}   
\end{figure*}


\section{Proof of Theorem 4}
\IEEEproof
Define the state that UAV ad-hoc netwok has the latest two strategy profiles $s(t-1)$ and $s(t)$ to be tuple $a(t)=[s(t-1),s(t),x(t)]$, where $x(t)=[x_1(t),x_2(t),...,x_M(t)]$.
\par Let $\epsilon=e^{-\frac{1}{\tau}}$. When $\epsilon$ approaches to $0+$, $\omega=0$, then the unperturbed process equals to UAVs never altering strategies. It is obvious that the recurrent classes of UAV ad-hoc network game with unperturbed process are $[s,s,\textbf{0}]$, where two strategy profiles are identical and any element in $x(t)$ is $0$. Any transition in the process of SPBLLA can be written as $$z^1=[s^0,s_1,x_1]\rightarrow z^2[s^1,s_2,x_2],$$ to be specific$$x_i^1=0 \Rightarrow \left\{
            \begin{array}{rcl}
x_i^2=0, s_i^2=s_i^1, or \\
x_i^2=1, s_i^2\in C_i(s_i^1),\\
            \end{array} \right. $$
            $$ x_i^1=1 \Rightarrow x_i^2=0, s_i^2\in {s_i^0,s_i^1}.$$
The probability of transition from $a^1$ to $a^2$ is
\begin{small}
\begin{equation}
\begin{split}
P_{\epsilon a^1\rightarrow a^2}=\Bigg(\prod_{i:x_i^1=0,x_i^2=0}(1-\omega)\Bigg)\Bigg(\prod_{i:x_i^1=0,x_i^2=1}\frac{\omega}{|C_i(s_i^1)|}\Bigg)
\\ \Bigg(\prod_{i:x_i^i=1,s_i^2=s_i^0}\frac{\epsilon^{-U_i(s^0)}}{\epsilon^{-U_i(s^0)}+\epsilon^{-U_i(s^1)}}\Bigg)
\\
 \Bigg(\prod_{i:x_i^i=1,s_i^2=s_i^1}\frac{\epsilon^{-U_i(s^1)}}{\epsilon^{-U_i(s^0)}+\epsilon^{-U_i(s^1)}}\Bigg).
\end{split}
\end{equation}
\end{small}
Denote
\begin{equation}
\begin{split}
V_i(s^0,s^1)&:=max\{U_i(s^0),U_i(s^1)\},
\end{split}
\end{equation}
\begin{equation}
\begin{split}
s(x)&:=\sum_ix_i.
\end{split}
\end{equation}
Multiplying the numerator and denominator of $P_{\epsilon z^1\rightarrow z^2}$ by $\prod_{i\in M}\epsilon^{V_i(s^0,s^1)}$, and

let $$R(a^1\rightarrow a^2)=\epsilon_{ms(x^2)}+$$
$$
\sum_{i:x_i^i=1,s_i^2=s_i^0}(V_i(s^0,s^1)-U_i(s^0))+$$
$$
\sum_{i:x_i^i=1,s_i^2=s_i^1}(V_i(s^0,s^1)-U_i(s^1)).$$
\begin{small}
\begin{equation}
\begin{split}
\frac{P_{\epsilon a^1\rightarrow a^2}}{\epsilon^{R(z^1\rightarrow z^2)}}=\Bigg(\prod_{i:x_i^1=0,x_i^2=0}(1-\epsilon^m)\Bigg)\Bigg(\prod_{i:x_i^1=0,x_i^2=1}\frac{1}{|C_i(s_i^1)|}\Bigg)
\\ \Bigg(\prod_{i:x_i^i=1,s_i^2=s_i^0}\frac{1}{\epsilon^{V_i(s^0,s^1)-U_i(s^0)}+\epsilon^{V_i(s^0,s^1)-U_i(s^1)}}\Bigg)
\\ \Bigg(\prod_{i:x_i^i=1,s_i^2=s_i^1}\frac{1}{\epsilon^{V_i(s^0,s^1)-U_i(s^0)}+\epsilon^{V_i(s^0,s^1)-U_i(s^1)}}\Bigg).
\end{split}
\end{equation}
\end{small}
\par Since when $\epsilon\rightarrow0+$, $$\epsilon^{V_i(s^0,s^1)-U_i(s^0)}+\epsilon^{V_i(s^0,s^1)-U_i(s^1)}\rightarrow1.$$
Thus, $0<lim_{\epsilon\rightarrow0}\frac{P_{\epsilon z^1\rightarrow z^2}}{\epsilon^{R(z^1\rightarrow z^2)}}<\infty$ and $R(z^1\rightarrow z^2)$ is the resistance of transition from $a^1$ to $a^2$. Then SPBLLA induces a regular perturbed markov process.

\par According to Lemma 1 in \cite{eco}, we learn that the stochastically stable strategy profiles are strategy profiles with minimum stochastic potential, and a strategy profile is a stochastically stable strategy profile only if it is in a recurrent class of unperturbed process $P^0$. Then we only need to prove that the root of minimum trees of strategy profiles $S$ is the maximizer of $\phi$. Denote $3$ kinds of strategy profile $X$, $Y$ and $Z$ ,where $X$ is what cannot be written as $[s,s,\textbf{0}]$, $Y=[s,s,\textbf{0}]$ is what is not a maximizer of $\phi$, and $Z=[s,s,\textbf{0}]$ is what is a maximizer of $\phi$. Only $Y$ and $Z$ are the candidates for stochastically stable strategy profiles.

\par Due to $X$ cannot be the stochastically stable strategy profile, we focus on $Y$ and $Z$. Build a minimum resistance tree $T$ rooted at $s*$, and some edges have multiple UAVs alter strategies. At each edge, there are several UAVs such that $s_i^k\neq s_i^{k-1}$, which compose a group $G\subseteq M$. Then the probability of transition from $s$ to $s'$ is
$$P_{\epsilon s\rightarrow s'}=\sum_{S\subseteq M: G\subseteq S}\epsilon^{m|S|}(1-\epsilon^m)^{|M\backslash S|}\prod _{i\in M}\frac{\epsilon^{-U_i(s')}}{\epsilon^{-U_i(s')}+\epsilon^{-U_i(s)}},$$
the resistance of transition from $s$ to $s'$ is
\begin{equation}
R(s\rightarrow s')=m|G|+\sum_{i\in G}(V_i(s,s')-U_i(s')).
\end{equation}

\par Denote the upper bound of $V_i(s,s')-U_i(s')$ for any $i\in M$ to be $\Delta$, the resistance of the transition from $s$ to $s'$ satisfies the inequality as follows:
\begin{equation}
m|G|+\Delta |G|\geq R(s\rightarrow s')\geq m|G|.
\label{1}
\end{equation}

\par In the first condition, there is merely a single UAV altering strategy for each edge $[s\rightarrow \tilde{s}]\in T$, then the argument in \cite{SPBLLA} illustrates that $s*$ is a maximizer of the potential function.

\par In another condition, assuming that there exists an edge $[s\rightarrow \tilde{s}]\in T$ with multiple UAVs atlering strategies, Let $\tilde{G}$ consist of the group of these multiple UAVs. The resistance of this edge meets
\begin{equation}
R(s\rightarrow \tilde{s})\geq m|\tilde{G}|.
\label{mG}
\end{equation}

\par Consider a path $L=\{s=s^0\rightarrow s^1\rightarrow...\rightarrow s^{\ \tilde{G}|}\}$, where for any $k\in {1,...,|\tilde{G}|}$, there is only one UAV altering strategy and the rest UAVs keep the original strategies. equ. (\ref{1}) shows that the resistance of these edges is less than
\begin{equation}
R(s^{k-1}\rightarrow s^k)\leq m+ \Delta,
\end{equation}
and the resistance of $L$ is at most
\begin{equation}
R(L)\leq |\tilde{G}|(m+\Delta).
\end{equation}

\par Build a new tree $T'$ rooted at $s*$ as well but remove redundant edges of $L^R$ from $T$ and add edges of $L$ to $T$. The redundant edges consist of $[s\rightarrow \tilde{s}]$ and other directed edges rooted at strategy profiles in $\{s^1,s^2,...s^{|\tilde{G}|-1}\}$ but not in $L$. Fig.~\ref{fig:treeT'} presents how the new tree $T'$ forms. We can deduce from equ. (\ref{mG}) that $R(s\rightarrow \tilde{s})\geq m|\tilde{G}|$ and the remanding edges in $L^R$ is at least $m$. Then, the resistance of $L^R$ satisfies
\begin{equation}
R(P^R)\geq m|\tilde{G}|+m(|\tilde{G}|-1).
\end{equation}
The resistance of tree $T'$ is
\begin{small}
\begin{equation}
\begin{split}
R(T') & =R(T)+R(L)-R(P^R)
\\    &\leq R(T)+|\tilde{G}|(m+\Delta)-(m|\tilde{G}|+m(|\tilde{G}|-1))
\\    & =R(T)+\Delta|\tilde{G}|-m(|\tilde{G}|-1).
\end{split}
\end{equation}
\end{small}
Since $|\tilde{G}|\geq2$, if $m> 2\Delta$ then $R(T')<R(T)$. The conflict implies that trees with edges, which has multiple UAVs changing strategies, are not the minimum resistance trees and this condition do not exist. Therefore, only maximizers of potential function are stochastically stable.
\endIEEEproof

\section{Proof of Theorem 5}
\IEEEproof
 Notice that $\Delta_{max}=|U_i(s')-U_i(s)|_{max}$, where the distinction between $s'$ and $s$ is capable to be made in one iteration. Define $\delta(x)$ as the discrepancy of $x$ in an iteration.
In one iteration
\begin{small}
\begin{equation}
\begin{split}
&U_i(s')-U_i(s)
\\=&\Bigg(A\frac{E}{\sum\limits_{1\leq n\leq N}P'_{in}}+B\{\sum[(P'_i-\gamma\sigma_{ip}(s_{-i}))\otimes C_i]-\alpha\bar{D}'_i\}
\\&+C\frac{\tilde{D}'_i}{D}\Bigg)-\Bigg(A\frac{E}{\sum\limits_{1\leq n\leq N}P_{in}}+B\{\sum[(P_i-\gamma\sigma_{ip}(s_{-i}))\otimes C_i]
\\&
-\alpha\bar{D}_i\}+C\frac{\tilde{D}_i}{D}\Bigg)
\end{split}
\end{equation}
\begin{equation}
\begin{split}
\\
=&\delta(\frac{AE}{\sum\limits_{1\leq n\leq N}P_{in}}+B\sum\limits_{1\leq n\leq N}P_{in})+\delta(B\gamma\sum\sigma_{ip}(s_{-i}))
\\&
+\delta(-B\alpha D_i)+\delta(B\alpha\kappa\sum\sigma_{ia})+\delta(C\frac{\tilde{D}_i}{D}).
\end{split}
\end{equation}
\end{small}
\par If each $\delta(x)$ approaches maximum, $U_i(s')-U_i(s)$ gets the maximum. The maximum $\delta(x)$ are as follows.
\begin{small}
\begin{equation}
\begin{split}
&\delta(\frac{AE}{\sum\limits_{1\leq n\leq N}P'_{in}}+B\sum\limits_{1\leq n\leq N}P'_{in})_{max}
\\=&AE(\frac{1}{\sum C_i\cdot P_1}-\frac{1}{\sum C_i(P_1+\Delta P)})+B\sum C_i\cdot\Delta P
\\=&AE\frac{\Delta P}{\sum C_i\cdot P_1\cdot (P_1+\Delta P)}+B\sum C_i\cdot\Delta P,
\end{split}
\label{delta1}
\end{equation}
\end{small}
\begin{small}
\begin{equation}
\delta(B\gamma\sum\sigma_{ip}(s_{-i}))_{max}=B\gamma\Delta P\cdot\sum(\sum\limits_{j\neq i}C_j\otimes C_i),
\end{equation}
\end{small}
\begin{equation}
\begin{split}
\delta(-B\alpha D_i)_{max}=&B\alpha\pi tan^2\theta[h_{nh}^2-(h_{nh}-\Delta h)^2]
\\=&B\alpha\pi tan^2\theta(2h_{nh}\Delta h-\Delta h^2),
\end{split}
\end{equation}
\begin{equation}
\delta(B\alpha\kappa\sum\sigma_{ia})_{max}=B\alpha\kappa(M-1)\pi tan^2\theta(2h_{nh}\Delta h-\Delta h^2),
\end{equation}
\begin{equation}
\delta(C\frac{\tilde{D}_i}{D})_{max}=\frac{C}{D}\pi tan^2\theta[h_{nh}^{2\beta}-(h_{nh}-\Delta h)^{2\beta}].
\label{delta5}
\end{equation}
Thus, $\Delta_{max}$ equals to the summation of the right value of Eqs. (\ref{delta1})--(\ref{delta5}). When $m>\Delta_{max}$, $m$ satisfies the assumption that $m$ is large enough.
\endIEEEproof




\ifCLASSOPTIONcaptionsoff
  \newpage
\fi

%





\end{document}